# Black Phosphorus Plasmonics: Anisotropic Elliptical Propagation and Nonlocality-Induced Canalization

*D. Correas-Serrano[1,2], J. S. Gomez-Diaz[1], A. Alvarez Melcon[2], and Andrea Alù[1]\**

[1] Department of Electrical and Computer Engineering, The University of Texas at Austin, Austin, TX 78712, USA

[2]Universidad Politecnica de Cartagena, 30202 Cartagena, Murcia, Spain.

**Keywords:** Plasmonics, uniaxial, hyperbolic materials, black phosphorus, nonlocality.

**Abstract:** We investigate unusual surface plasmons polariton (SPP) propagation and light-matter interactions in ultrathin black phosphorus (BP) films, a 2D material that exhibits exotic electrical and physical properties due to its extremely anisotropic crystal structure. Recently, it has been speculated that the ultra-confined surface plasmons supported by BP may present various topologies of wave propagation bands, ranging from anisotropic elliptic to hyperbolic, across the mid- and near-infrared regions of the electromagnetic spectrum. By carefully analyzing the natural nonlocal anisotropic optical conductivity of BP, derived using the Kubo formalism and an effective low-energy Hamiltonian, we demonstrate here that the SPP wavenumber cutoff imposed by nonlocality prohibits that they acquire an arbitrary hyperbolic topology, forcing operation in the canalization regime. The resulting nonlocality-induced canalization presents interesting properties, as it is inherently broadband, enables large light-matter interactions in the very near field, and allows extreme device miniaturization. We also determine fundamental bounds to the confinement of BP plasmons, which are significantly weaker than for graphene, thus allowing a larger local density of states. Our results confirm the potential of BP as a promising reconfigurable plasmonic platform, with exciting applications, such as planar hyperlenses, optoelectronic components, imaging, and communication systems.

**Introduction**

The rise of graphene in the past decade [1]–[4] has triggered extensive research on other two-dimensional materials with properties able to surmount some of graphene's shortcomings. Great effort has been devoted to transition metal dichalcogenides (TMDs) such as $MoS_2$ and $WS_2$ [5]–[8] due to their direct bandgap, an attractive feature to achieve high on-off ratios, which has proven elusive in graphene devices [9]. However, the low mobility in these materials (orders of magnitude below the one of graphene) has hindered their applicability. More recently, black phosphorus (BP) thin films, or phosphorene, has been rediscovered as a material able to simultaneously exhibit high on-off ratios and carrier mobility comparable to that of graphene [10]. Importantly, it may have yet untapped potential for mid-IR nano-optics, since an adequate selection of the film thickness results in a small bandgap and good optical transitions for realistic doping levels [11]. In this direction, perhaps the most striking feature of BP compared to other 2D materials is the naturally high in-plane anisotropy of its macroscopic physical properties, which arises due to its peculiar orthorhombic lattice with two clearly distinguishable axes, as depicted in Fig. 1.

BP has been successfully isolated as a monolayer and few-layer form by means of mechanical exfoliation and plasma thinning [12]. In addition, exciting possibilities to control the optical properties of BP have been demonstrated using different methods. For instance, Ref. [12] experimentally demonstrated high-resolution control of these properties by laser pruning, reducing thick BP films to few-layer forms and enabling local and straightforward engineering of the bandgap and the optical conductivity. Strong normal fields induced by dopants were used in [13] to experimentally demonstrate drastic tuning of the bandgap due to giant Stark effect, allowing to efficiently and swiftly switch the nature of BP from a moderate-bandgap semiconductor to a band-inverted semimetal. In the critical point before band inversion, BP becomes an anisotropic Dirac semimetal similar to graphene *only* in one direction, providing unprecedented design flexibility that may open venues for novel functionalities in planar optics. Importantly, this behavior can be obtained through an externally applied gate bias similarly to graphene devices. Mechanical strain has been predicted to enable extensive reconfiguration capabilities as well [14], e.g. by allowing to swap the higher-conductivity direction with a moderate biaxial or uniaxial strain [15].

Until now, most studies on the applied physics of BP have focused on its DC properties, aiming to realize fast and efficient electronic transistors and devices. Even though interesting optical responses, such as strong layer-dependent photoluminescence have been reported [16] – similarly to the case of $MoS_2$ [17] –, they have been mainly applied to confirm the direct and layer-sensitive bandgap of the material, rather than to the design of practical optoelectronic devices. One exciting direction to realize such components would be to exploit the extremely-confined anisotropic surface plasmons polaritons (SPPs) that BP supports at mid-IR owing to its inductive nature [18]-[19]. It is important to note that graphene cannot support low-loss plasmons at such high frequencies, due to the strong coupling of SPPs to optical phonons [3]. Localized plasmons in patterned BP films have also been theoretically studied in the context of plane wave transmission/reflection, showing the expected polarization dependence due to the material anisotropy [20]. Unfortunately, to our knowledge the few existing works on BP plasmonics have all considered a local description of the optical conductivity, so the fundamental performance bounds and potential practical applications of BP are not yet fully understood.

In this context, this paper presents a comprehensive study of BP plasmonic properties from terahertz to infrared (IR), unveiling the potential of this material as a platform for future planar and reconfigurable mid-IR devices with improved functionalities over alternative technologies. The main contributions of this paper are: (i) a clear discussion, using realistic material parameters, of the different SPPs topologies supported in BP films when taking into account both intraband and interband contributions; and (ii) the study of nonlocal effects in BP plasmons, shedding light into fundamental bounds on field confinement while also reporting unusual light-matter interactions and quantum nonlocality-related phenomena. Specifically, our study demonstrates that the hyperbolic response predicted by local conductivity models [19] cannot be realized in practice, and instead becomes a broadband, nonlocality-induced and deeply subwavelength canalization regime that can be directly applied to realize in-plane hyperlenses.

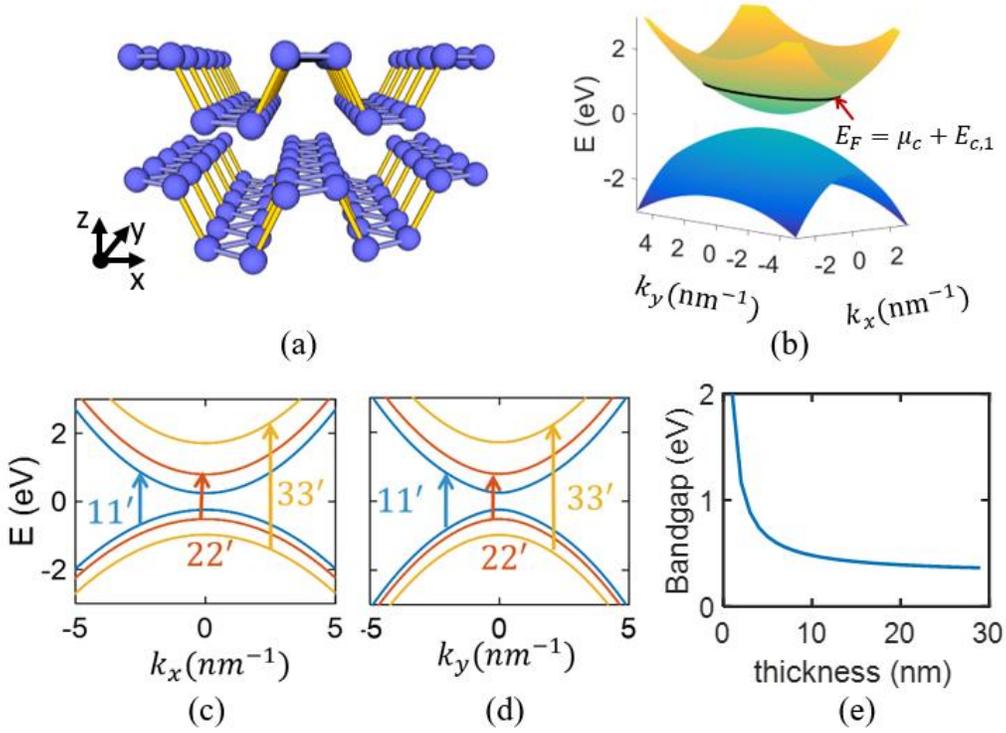

**Figure 1. Optical properties of black phosphorus.** (a) Lattice structure of a two-layer ultrathin film. Different colors are employed for visual clarity. (b) Energy-wavenumber dispersion diagram of electrons in BP. $E_F$, $E_{c,1}$, and $\mu_c$ are the Fermi and first-conduction band energies and the chemical potential, respectively. (c)-(d) First three quantized valence/conduction subbands along $k_x$ and $k_y$ and associated interband transition processes. (e) BP bandgap dependence versus layer thickness.

**Optical conductivity of BP thin films**

Throughout this paper we assume infinitesimally-thin layers of BP, which can be modeled by the space- and time-dispersive conductivity tensor

$$\bar{\bar{\sigma}}(\omega,\boldsymbol{q}) = \begin{pmatrix} \sigma_{xx}(\omega,\boldsymbol{q}) & \sigma_{xy}(\omega,\boldsymbol{q}) \\ \sigma_{yx}(\omega,\boldsymbol{q}) & \sigma_{yy}(\omega,\boldsymbol{q}) \end{pmatrix}, \qquad (1)$$

where $\omega = 2\pi f$ is the angular frequency, $\boldsymbol{q}$ is the wavevector, and the different components can be computed through the well-known Kubo formalism. Here we use the approximate two-dimensional Hamiltonian developed in [14] to model the in-plane electron dispersion, which is based on $k \cdot p$ theory and is valid near the $\Gamma$ point (low to moderate electron energies). We reproduce it here for completeness. It reads

$$\mathcal{H} = \begin{pmatrix} E_c + \eta_c k_x^2 + \nu_c k_y^2 & \gamma k_x + \beta k_y^2 \\ \gamma k_x + \beta k_y^2 & E_v - \eta_v k_x^2 - \nu_v k_y^2 \end{pmatrix}, \quad (2)$$

where $E_c$ and $E_v$ are the energies of the conduction and valence band edges, respectively, and $\eta_{c,v}$ and $\nu_{c,v}$ are related to the in-plane effective masses. For multilayer films, the above Hamiltonian describes electron dispersion in each layer. In this case, different valence and conduction subband energies must be considered for each layer $j$, with $E_{c,v}$ being replaced by $E_{(c,v),j}$, the edge energy values of the corresponding subband pairs. By choosing the previous values to match experimental data for the effective masses and bandgap of few-layer and bulk BP, we are able to accurately model its conductivity. We use the same values as in [21], which are consistent with experimental reports. This model allows to phenomenologically take into account bandgap shrinkage and widening due to giant Stark effect [13], [22] or mechanical strain, among others. The Kubo formula can be used calculate all the components of the conductivity tensor as a function of frequency and the wavevector as

$$\sigma_{\alpha\beta}(\omega, \mathbf{q}) = -\frac{i\hbar e^2}{2\pi^2} \sum_{ss'jj'} \int \frac{f(E_{sjk}) - f(E_{s'j'k'})}{E_{sjk} - E_{s'j'k'}} \frac{\langle \Phi_{sjk} | \hat{v}_\alpha | \Phi_{s'j'k'} \rangle \langle \Phi_{s'j'k'} | \hat{v}_\beta | \Phi_{sjk} \rangle}{E_{sjk} - E_{s'j'k'} + \hbar\omega + i\eta} d\mathbf{k} \quad (3)$$

In this equation, $\mathbf{k'} = \mathbf{k} + \mathbf{q}$, $\hbar$ is the reduced Planck constant, $e$ is the electron charge, $\eta$ is the finite damping, $f(\ldots)$ is the Fermi-Dirac distribution function, $\hat{v}_\alpha = \hbar^{-1} \partial_{k_\alpha} \mathcal{H}$ is the velocity operator, $E_{sjk}$ and $\Phi_{sjk}$ are the eigenvalues and eigenfunctions of $\mathcal{H}$, and the indices $ss' = \pm 1$ and $jj'$ denote the conduction/valence band and the subband indices, respectively. Due to the nonexistent overlap between different subbands in this model [14], [18], optical intra/interband transitions are only allowed between equal subbands, i.e. $j = j'$, as illustrated in Fig. 1c-d (otherwise the matrix elements in the Kubo integral vanish). Lastly, we define the chemical potential $\mu_c$ as the difference between the Fermi level and the first conduction band, i.e., $E_F = E_{c,1} + \mu_c$, and use a scattering rate of 3 meV.

In the local case, i.e., when $\mathbf{q} \to \mathbf{0}$, the optical conductivity tensor is diagonal and independent from the EM wavevector, yielding the commonly used local tensor

$$\bar{\bar{\sigma}}(\omega) = \begin{pmatrix} \sigma_{xx}(\omega) & 0 \\ 0 & \sigma_{yy}(\omega) \end{pmatrix}. \quad (4)$$

Figs. 2a-b illustrate the imaginary and real pats of the local conductivity ($\mathbf{q} \to 0$) of a 10 nm thick BP multilayer versus frequency for various values of chemical potential $\mu_c$. For this thickness, the bandgap is estimated to be 0.48 eV. Results confirm a Drude-like dispersion below the interband transition threshold and a high degree of anisotropy with ample tunability for small variations of the chemical potential. The $y$-component of the conductivity presents metallic behavior (inductive, i.e., Im[$\sigma$] > 0 under a $e^{i\omega t}$ time convention) for a wider range of frequencies than the $x$-component, which is the high-conductivity direction and shows a very distinct transition from metallic to dielectric (capacitive, i.e., Im[$\sigma$] < 0) response due to interband transitions. Interband processes become significant for frequencies larger than $2\pi f_0 \hbar \approx E_g + 2\mu_c$, where $E_g$ is the

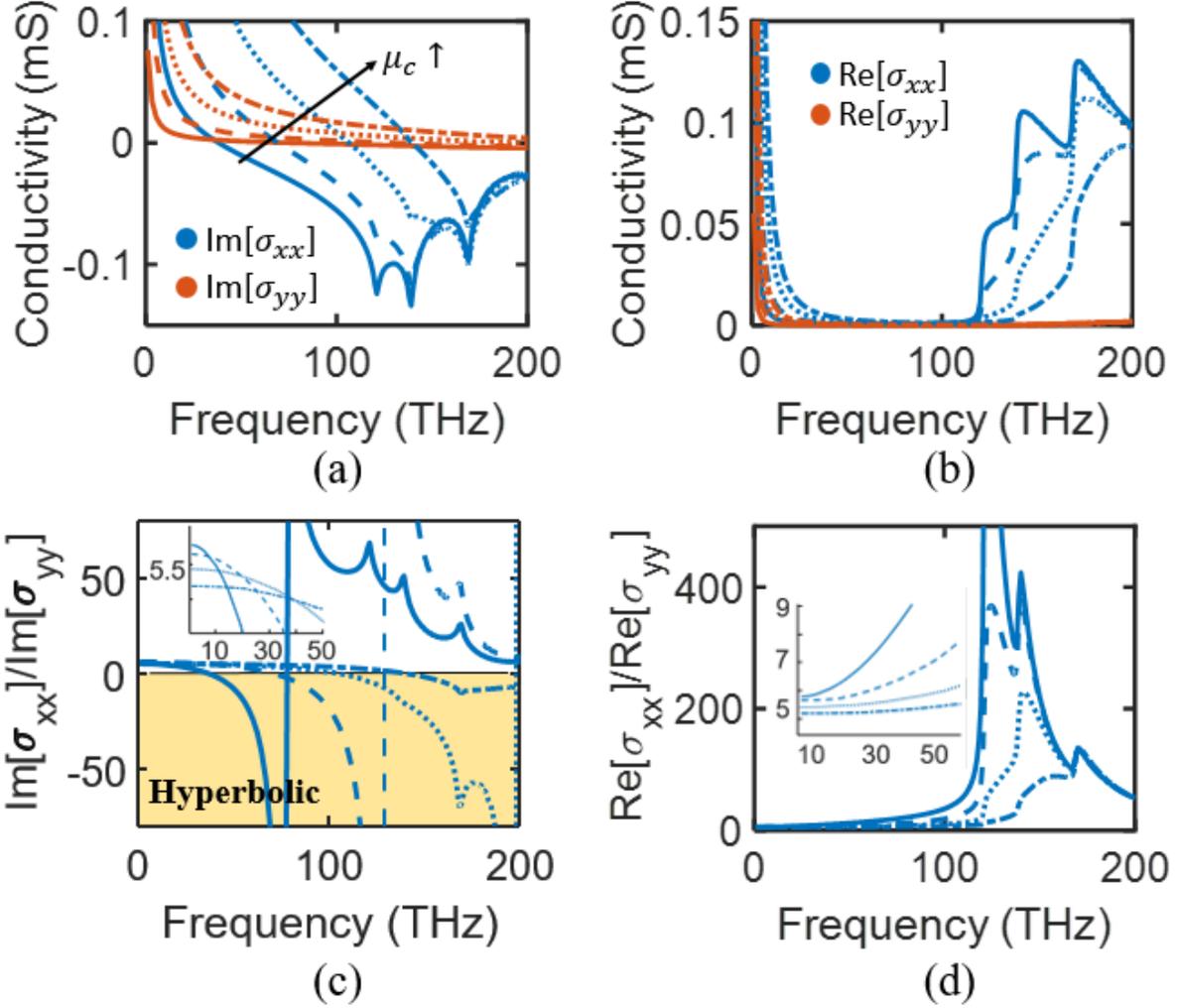

**Figure 2. Dispersion of BP optical conductivity.** Imaginary (a) and real (b) components of BP conductivity versus frequency plotted for various values of chemical potential. Different curves correspond to growing (as indicated by the arrow) values of chemical potential: 0.005 eV, 0.05 eV, 0.1 and 0.2 eV. Blue and red lines corresponds to $\sigma_{xx}$ and $\sigma_{yy}$ conductivity components, respectively. (c), (d) Anisotropy ratio of the imaginary and real parts of the conductivity (insets show detail at low frequencies). Hyperbolic regime appears at the frequencies where the ratio $\text{Im}[\sigma_{xx}]/\text{Im}[\sigma_{yy}]$ is negative (shaded region in panel c). Other parameters are similar to Fig. 1.

bandgap energy. Due to Pauli blocking, these frequency points are separated by approximately $\Delta f \approx \Delta \mu_c / \pi \hbar$ and can therefore be controlled in practice by modifying BP chemical potential through an externally applied electrostatic bias [18]. The peaks observed in the conductivity above this frequency appear due to the different interband transition frequencies of the subband pairs associated to each layer (see Fig. 1c-d).

Figs 2c-d explicitly show a large degree of anisotropy in the real and imaginary parts of the conductivity components. In the Drude regime, we find that the ratios $\text{Im}[\sigma_{xx}]/(\text{Im}[\sigma_{yy}]$ and $(\text{Re}[\sigma_{xx}]/\text{Re}[\sigma_{yy}]$ are weakly dependent on $\mu_c$, and hover around 5. This anisotropy becomes strongly dependent on chemical potential as we approach the interband threshold for the *x*-component, which marks the beginning of the hyperbolic regime predicted in [19]. Before this point, the anisotropy in the imaginary part continuously decreases due to the interband processes counteracting the inductive intraband contributions, while loss monotonically increases. This ratio evidently crosses unity, allowing isotropic phase propagation with anisotropic plasmon decay, something unique to BP. The hyperbolic regime has a broad bandwidth, until $\sigma_{yy}$ also becomes dielectric. After this point, BP behaves as a thin dielectric layer with very large anisotropy.

Overall, this study demonstrates the dispersive, tunable and extremely anisotropic electromagnetic response of BP. This response is indeed very rich and includes anisotropic elliptic quasi-transverse magnetic (TM, i.e., $\text{Im}[\sigma_{xx}] > 0, \text{Im}[\sigma_{yy}] > 0$ ) and quasi-transverse electric (TE, i.e., $\text{Im}[\sigma_{xx}] < 0, \text{Im}[\sigma_{yy}] < 0$ ) behaviors at low and very high frequencies, respectively, as well as an intrinsic hyperbolic frequency band ( $\text{Im}[\sigma_{xx}] < 0, \text{Im}[\sigma_{yy}] > 0$ ) and two clearly-defined topological transitions that implement *σ*-near-zero topologies. To our knowledge, hyperbolic plasmons have only been previously reported at THz and far-IR frequencies using nanostructured graphene [23], [24] and at optics by patterning silver [25] and therefore BP may potentially fill this frequency gap by enabling SPPs with hyperbolic topologies at mid-IR. However, as detailed later in the text, nonlocal effects may play a dominant role in such scenarios, dramatically modifying the expected hyperbolic response of BP.

**Anisotropic surface plasmons in BP thin films**

In this section, we solve the exact dispersion relation of surface waves in BP and compute the associated fields, in order to quantitatively characterize the operation regime of the supported SPPs discussed earlier. For now, we still consider a local description of the conductivity. The dispersion of the supported plasmons is given by [24], [26], [27]

$$k_0 q_z \left(4+\eta_0^2 \left(\sigma_{xx}\sigma_{yy}-\sigma_{xy}\sigma_{yx}\right)\right)-2\eta_0 k_0^2\left(\sigma_{xx}+\sigma_{yy}\right)+2\eta_0\left(q_x^2\sigma_{xx}+\sigma_{yy}q_y^2+q_x q_y \sigma_{xy}\sigma_{yx}\right)=0 \quad (5)$$

where $k_0^2=q_x^2+q_y^2+q_z^2$, $\eta_0$ is the free-space impedance, and the possible wavenumber-dependence of the conductivity components has been omitted. This equation can be numerically solved with conventional root-finding algorithms in the complex plane [24]. Fig. 3 illustrates the three main topologies discussed in the previous section, which can be found in BP at different operation frequencies. For each topology, the figure shows the associated isofrequency contour (IFC) and the normal component of the electric field along the BP excited by a *z*-oriented dipole located 10 nm above the layer. Fig. 3a corresponds to the elliptical regime that is found at "low" frequencies (50 THz) where the intraband contributions dominate. The degree of anisotropy $\text{Im}[\sigma_{xx}]/\text{Im}[\sigma_{yy}]$ is roughly 5 in this scenario, which results into a significant canalization along the *x*-axis (armchair direction). Note that in this Drude regime, i.e., when interband transitions are negligible, the ratio between the axial inductances is weakly dependent on frequency and chemical potential (see Fig. 2), so this type of topology is the most common at low frequencies.

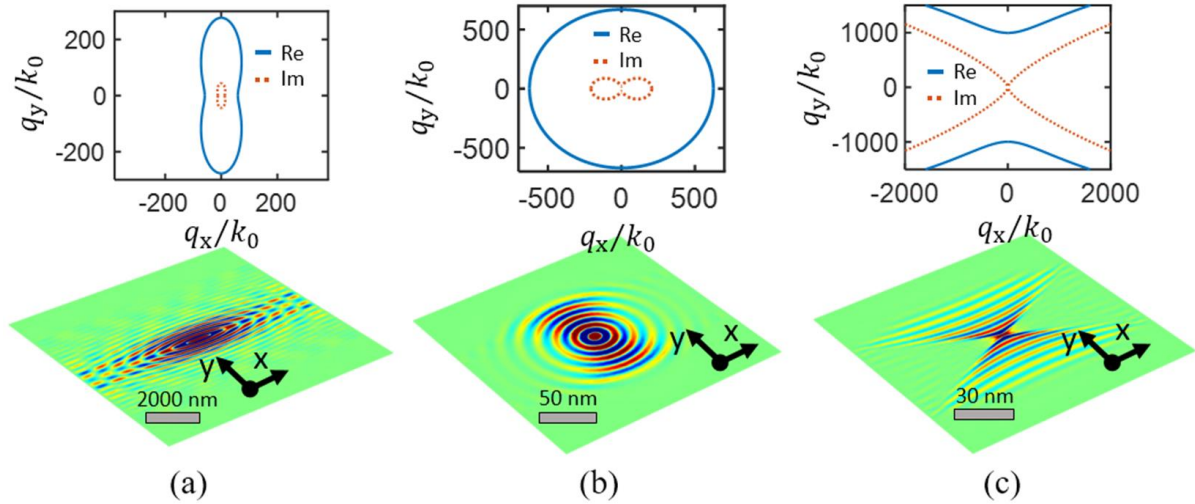

**Figure 3. Anisotropic surface plasmons supported by black phosphorus.** Isofrequency contours are computed using Eq. (3). Color maps show the *z*-component of the electric field excited by a dipole located 10 nm above the surface. (a) Anisotropic elliptic plasmons, operation frequency is $f = 50$ THz, chemical potential $\mu_c = 0.1$ eV (b) Interband-induced quasi-isotropic plasmons, $f = 100$ THz, $\mu_c = 0.1$ eV (c) Hyperbolic plasmons, $f = 80$ THz, $\mu_c = 0.05$ eV.

The middle panel corresponds to the quasi-isotropic point for $\mu_c = 0.1$ eV, which occurs around 100 THz. As anticipated in the previous section, phase propagation is uniform, but loss is clearly higher along the *x*-axis due to interband transitions. This unusual operation regime may be useful to realize polarization-sensitive absorbers. We note that around this frequency the conductivity is very sensitive to changes in doping. Lastly, the right panel illustrates the natural hyperbolic regime for $\mu_c = 0.05$ eV at 80 THz. We do stress that local, homogeneous BP is able to intrinsically support hyperbolic plasmons thanks to its inherent anisotropy, and without requiring complicated patterning process as the case of graphene [23], [24] and silver [25] A remarkable property of the latter two examples is the extremely large wavenumbers of the supported plasmons, with wavelengths hundreds of times smaller than free-space. Properly designed, such wave-confinement can potentially lead to ultra-miniaturized devices. We have found that this response is always present in BP films operating near interband frequencies, as a result of $\sigma_{yy}$ always being very small when $\sigma_{xx}$ approaches the transition from metallic to dielectric. This can be mitigated only partially by modifying the doping or film thickness. Essentially, we have found that it is impossible to obtain hyperbolic responses with wavenumbers significantly lower than the ones considered in Fig. 3. It is well known that extremely confined surface waves are associated with a drastic enhancement of the local fields and the spontaneous emission rate of nearby sources, both of which are desired features in applications such as hyperlensing, nonlinear optics, or heat transfer [28]. However, several factors limit the response of any plasmonic device, and values as large as those reported here may lead to important practical difficulties or unexpected effects. These include achieving unwanted coupling to sources and detectors, increased loss due to impurities, and spatial dispersion phenomena. In- and out-coupling of energy to BP films can be done with near-field techniques similar to those used to excite plasmons in graphene [29], and its study is out of the scope of this paper. Loss mechanisms can be phenomenologically added to the model through the parameter $\eta$ in Eq. (3) with predictable results. On the other hand, spatial dispersion has less evident effects that depend on the electron dispersion in the lattice even in the absence of impurities, and as such they are not expected to be mitigated by improved fabrication processes. Lastly, note that due to the conductivity model used, we have neglected possible finite-thickness effects.

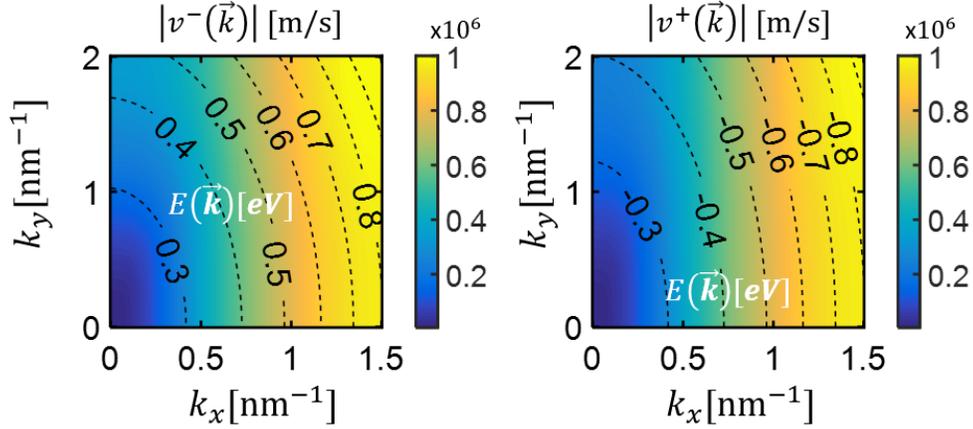

**Figure 4.** Contour plots of the electron energies in the first conduction band, overlaid with a colormap of the electron velocity.

**Spatial dispersion in realistic BP thin films**

Results in the previous section evidenced that the most exotic operation regimes in BP thin films involve extremely confined plasmons. Thus, the common assumption of a local conductivity, i.e. $q \to 0$ in Eq. (3) may not be valid. In the following we explicitly consider the wavevector-dependence of the conductivity to more accurately predict the topology of BP plasmons by assuming $q \neq 0$ in the Kubo formula and finding the fully populated conductivity tensor for any electromagnetic wavevector. Although computationally cumbersome, this approach is expected to be very accurate and to yield reliable insights into the performance limits of BP plasmons. We expect simplified models to be developed in the near future, for instance by using a phenomenological approach to introduce nonlocal corrections to the local Drude model within the Thomas-Fermi approximation. Similar models have been used in the study of traditional metallic nanostructures such as spheres and dimers [30] and we recently outlined how this nonlocal description of 3D media may be adapted to ultrathin materials [31].

In this context, it is illustrative to discuss the differences between BP and graphene with respect to nonlocal response. The simplest way to qualitative understand the influence of this phenomena is in terms of the electron velocity, with the so-called Fermi velocity $v_F$ corresponding to electrons with energies near the Fermi level. In general, the electron velocity is given by

$$v^\pm = \frac{1}{\hbar}\nabla_k E^\pm(k), \qquad (6)$$

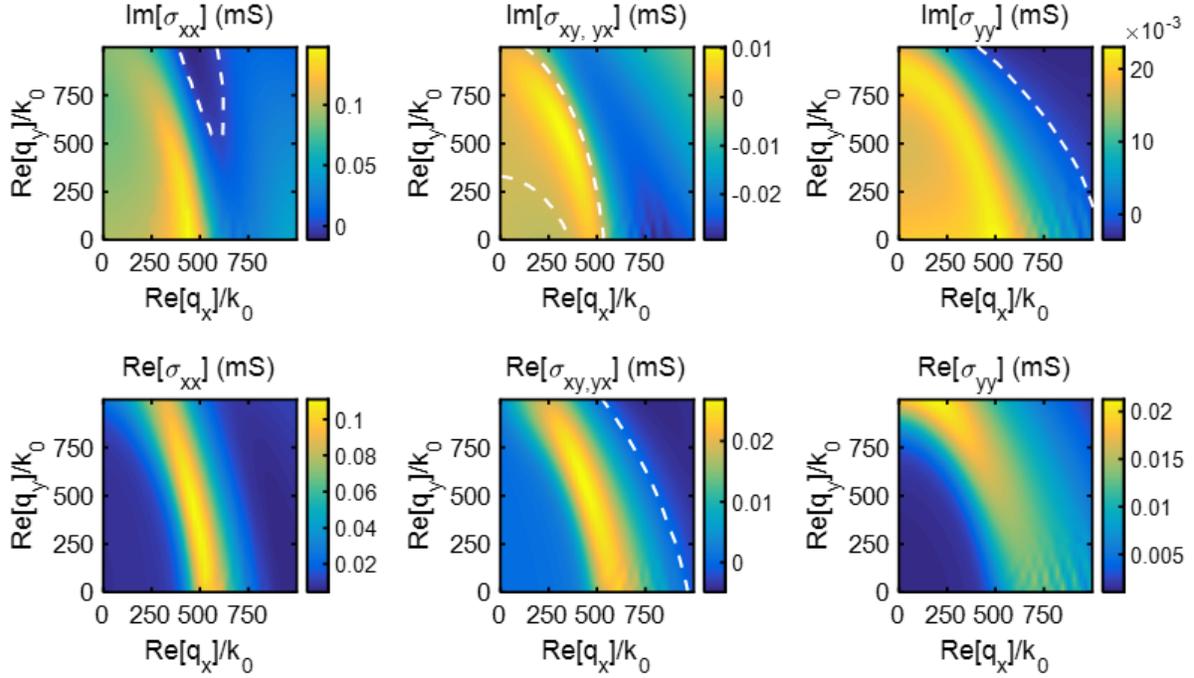

**Figure 5.** Nonlocal conductivity components of black phosphorus versus real in-plane wavenumbers $q_x/k_0$ and $q_y/k_0$, computed for $\mu_c = 0.1$ eV at 50 THz. White dashed lines indicate zero-crossings

where $E^{\pm}(k)$ is the electron dispersion relation, given by the eigenvalues of the Hamiltonian. In graphene, $E^{\pm}(k) = \pm\hbar v_F|k|$ [2] is the well-known linear dispersion law near the Dirac points, and leads to an energy-independent $v^{\pm} = v_F$. As shown in [31], this implies that the magnitude of the wavevector of propagating plasmons is necessarily bounded by $|k| < c/v_F k_0 \approx 300 k_0$ for any device based on graphene [31].

In BP, however, the picture is more complicated due to anisotropy, the non-symmetric conduction and valence bands, and the energy-dependent electron velocities as a result of the quadratic $k$-terms in the Hamiltonian. Analytical expressions can be obtained for $v^{\pm}$ by applying Eq. (6) to the eigenvalues of the Hamiltonian, but they involve cumbersome algebra and are omitted here for the sake of clarity. The color maps in Fig. 4 show the velocity in the conduction (left) and valence (right) bands in the vicinity of the Γ point. The contours represent the electron energies. We observe velocity values range from roughly $10^5$ m/s for the lowest energy electrons to around $10^6$ for energies around 1 eV (recall that $v_F = 10^6$ m/s = $c/300$ in graphene). Interestingly, this suggests that (i) plasmonic devices based on BP may support larger wavenumbers than their graphene counterparts, and (ii) the bounds may be doping dependent.

Fig. 5 shows the nonlocal conductivity components for $f = 50$ THz and $\mu_c = 0.1$ eV for a wide range of in-plane wavenumbers $q_x$ and $q_y$. This scenario corresponds to the elliptic regime illustrated in Fig. 3a. For small wavenumbers, these values converge to the local conductivity, with $\sigma_{xy} = \sigma_{yx} = 0$. As $\boldsymbol{q}$ increases, so do the non-diagonal terms and the real parts of the diagonal components. This implies that large wavenumbers force a non-negligible in-plane rotation of power from the longitudinal to the transverse direction, similarly to graphene [31] and other plasmonic systems [30]. Note, however, that this is not a dissipative process. In addition, very large $q$ may cause the conductivity components to change from metallic to dielectric or vice versa (white dashed lines show the zero-crossings). In this example the transition from metallic to dielectric response is different for $\sigma_{xx}$ and $\sigma_{yy}$, but in both cases occurs for $q \approx 600 k_0$. As expected from earlier discussions based on $v^{\pm}$, this value is significantly larger than in graphene, where $|q| < 300 k_0$ [31].

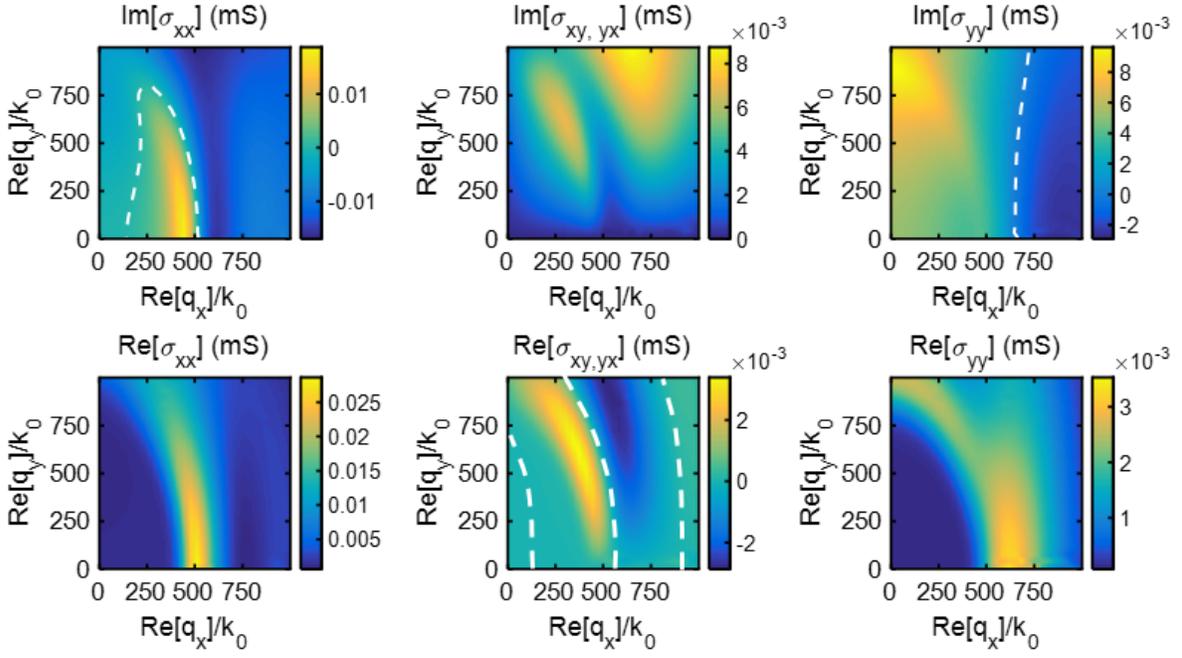

**Figure 6.** Nonlocal conductivity components of black phosphorus versus real in-plane wavenumbers $q_x/k_0$ and $q_y/k_0$, at operation frequency of 80 THz and $\mu_c = 0.05$ eV. While dashed lines indicate zero-crossings

Fig. 6 shows the nonlocal conductivity components corresponding to the hyperbolic regime previously considered (Fig. 3c), i.e., $f = 80$ THz and $\mu_c = 0.05$ eV. In this case, the local model predicts capacitive $\sigma_{xx}$ and metallic $\sigma_{yy}$, hence the hyperbolic response. However, it is clear that nonlocality causes $\sigma_{xx}$ to become metallic for a wide range of $q_x$, potentially changing the plasmon dispersion, as we will see later. Regarding $\sigma_{yy}$, it once again remains metallic until $|q| \approx 600 k_0$.

Let us focus again on the surface waves supported by the BP films under consideration, now taking its nonlocal response into account. Fig. 7 shows the IFCs computed with Eq. (3) assuming the fully populated nonlocal $\bar{\bar{\sigma}}$, with each $\sigma_{\alpha\beta}$ being replaced by $\sigma_{\alpha\beta}(q_x, q_y)$. Panels (a), (c) correspond to the case of elliptical plasmons (Fig. 5). Here, nonlocality has a moderate effect on the IFCs due to the relatively low wavenumbers involved. Interestingly, spatial dispersion *increases* the confinement along the *y*-direction, something that does not happen in graphene plasmons along infinite sheets [32]. This is in agreement with Figs. 5a-5c, which show a slight reduction of Im$[\sigma_{yy}]$ and Im$[\sigma_{xx}]$ when $q_y$ increases from zero up to $q_y \approx 500k_0$. The increased inductance makes plasmons more confined. Plasmons along the *x* direction are barely affected, and overall loss increases for all directions.

Panels (b), (d) show the IFCs of the hyperbolic case, related to Fig. 6. The effect of nonlocality is much more pronounced here, with the hyperbolic dispersion being replaced by an almost perfect canalization along *y*. In order to understand this phenomenon, it is helpful to once again consider the corresponding conductivity plots in Fig. 6. First, it is clear that increasing $q_y$ for low $q_x$ results in *higher* Im$[\sigma_{yy}]$ (lower inductance, contrary to the elliptic scenario), which causes a significant shift of the vertex of the hyperbola to lower $q_y$. Near $q_x = 0$, the hyperbolic shape can still be noticed, but as $q_x$ increases the topology changes drastically and the IFC closes around $q_x \approx 400k_0$. This is related to the unusual nonlocality induced dielectric-metallic transition shown in Fig. 6a around $q_x \approx 200k_0$. We have found that, within this model, nonlocality always prevents plasmons from exhibiting a true hyperbolic topology for a significant range of wavenumbers, in contrast with previous reports that considered only local response and predicted hyperbolic plasmons [19]. We stress, however, that more complex structures such as arrays of strips or patches might still support hyperbolic plasmons similar to other nanostructured metasurfaces [23], [25] although they should be designed to operate with lower wavenumbers.

We emphasize that this study assumes an infinitesimally thin layer, i.e., we are assuming a certain wavevector on the surface studying the material response, and therefore we do not consider possible bulk modes supported by the actual BP thin film. Within this assumption, our results aimed to probe the quantum-induced limits of BP plasmonics associated to spatial dispersion and have further demonstrated the immense potential of this material for manipulating IR light at the nanoscale. Specifically, we remark the looser physical bounds in terms of field confinement and the unusual phenomena related to nonlocality, in stark contrast with graphene sheets, where spatial dispersion reduces field confinement and increases loss. We envision that these features may be exploited in the future as additional degrees of freedom in the design of deeply subwavelength plasmonic devices.

**Light-matter interactions in BP thin films**

An effective and simple metric of the magnitude of light-matter interactions enabled by a metasurface is given by the spontaneous emission rate (SER) of nearby sources [23],[24]. Therefore, it is of evident importance to understand how this metric compares to other similar technologies, and how spatial dispersion may affect it. It can be computed as

$$SER = \frac{P}{P_0} = 1 + \frac{6\pi}{|\vec{\mu}_p|k_0} \vec{\mu}_p \cdot \mathrm{Im}\left[\bar{\bar{G}}_S\left(\vec{r}_0, \vec{r}_0, \omega\right)\right] \cdot \vec{\mu}_p, \qquad (6)$$

where $P_0$ and $P$ are the power generated by the emitter when isolated in free-space and when located above the BP film; $\vec{\mu}_p$ and $\vec{r}_0$ denote the dipole orientation and position, $k_0$ is the free-space wavenumber and $\bar{\bar{G}}_S(\vec{r}_0, \vec{r}_0, \omega)$ is the scattered term of the Green's function of the entire

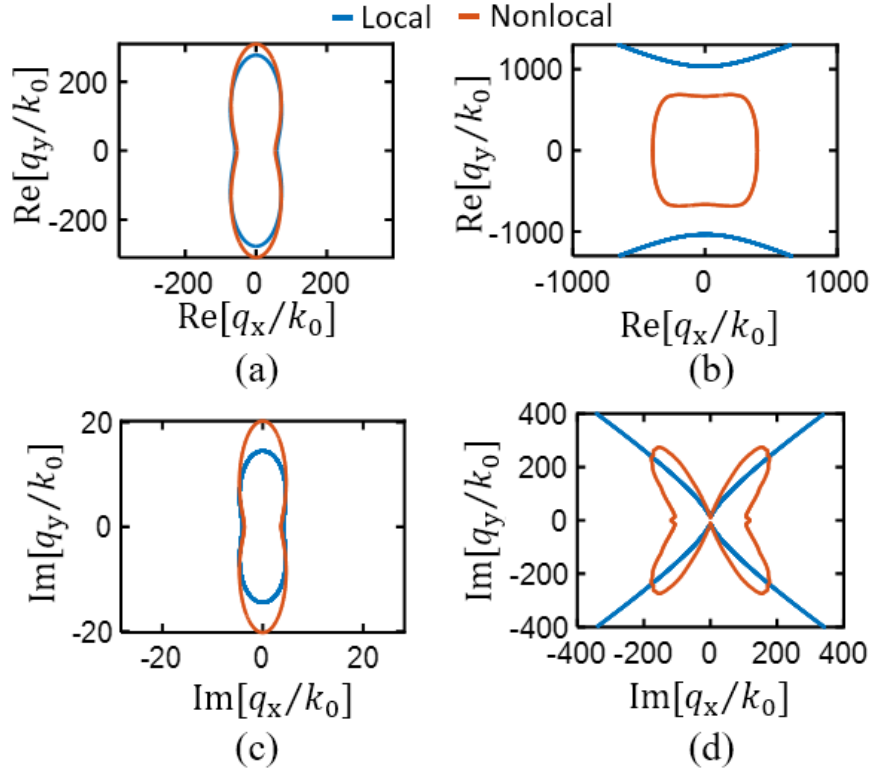

**Figure 7.** Influence of spatial dispersion in the isofrequency contours of SPP supported by black phosphorus films. Results are computed using Eq. (6), modelling BP optical conductivity with the Kubo formula (see Eq. (3)) with and without nonlocal effects taken into account. (a), (c) Operation frequency $f = 50$ THz, chemical potential $\mu_c = 0.1$ eV (same as Fig. 5). (b), (d) Operation frequency $f = 80$ THz, chemical potential $\mu_c = 0.05$ eV (same as Fig. 6).

system. In plasmonic structures, a drastic enhancement of power radiated by a nearby source is typically observed due to efficient coupling of the evanescent spectrum to surface waves with high $q$. Fig. 8 shows the SER of a $z$-oriented dipole versus its distance to the two BP films studied in the previous section, for the local and nonlocal models. We also plot the SER for a homogeneous graphene sheet with the same $\mu_c$ and $\eta$. Although this comparison is somehow arbitrary (many practical factors such as achievable scattering rate and ease of doping should be taken into account), it allows to easily compare results with a more familiar platform. Fig. 8a corresponds to the elliptical regime at 50 THz, with $\mu_c = 0.1$ eV. In this case the local model presents good accuracy due to the small effect that nonlocality has on plasmon dispersion. Since spatial dispersion slightly increases the confinement along the $y$-direction in this case (see Fig. 7a), the SER is in fact higher than predicted by the local model when the source is very close to the surface, and lower as it is moved farther. This occurs because high-$q$ waves decay faster in free-space. The SER associated to a homogeneous graphene sheet shows a similar trend versus distance, but there is a larger discrepancy between local and nonlocal models. This is an expected effect of spatial dispersion in both materials, as discussed in previous sections. Fig. 8b corresponds to the hyperbolic regime at 80 THz for $\mu_c = 0.05$ eV. For sources very close to the surface, the SER is larger than in Fig. 8a, but it decays more rapidly with distance due to the larger $q$ of the supported

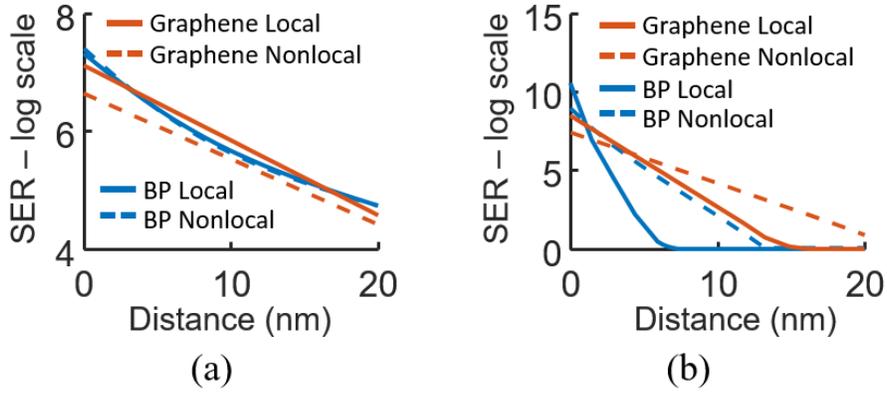

**Figure 8.** Spontaneous emission rate of a *z*-oriented point source versus its distance to BP thin films and graphene sheets, considering local and realistic nonlocal responses. Parameters are the same as (a) Fig. 7a ($f = 50$ THz, $\mu_c = 0.1$ eV) and (b) Fig. 7b ($f = 80$ THz, $\mu_c = 0.05$ eV)

plasmons (see Fig. 7b). The disagreement between nonlocal and local models is more pronounced in this case for the same reason. As a consequence, the energy coupled is orders of magnitude larger than predicted by the local mode when the source separation above a few nanometers. Overall, BP enables strong light-matter interactions, comparable to graphene, while providing additional freedom in tailoring plasmon dispersion and higher upper limits in the SER.

**Conclusion and Outlook**

We have investigated BP thin films as a potential IR plasmonic material with exotic optical properties, such as strong broadband anisotropy and a natural hyperbolic response. To this end, we have used a 2D description of its conductivity based on the Kubo formula and an approximate Hamiltonian. We have illustrated different plasmonic operation regimes, from elliptic to hyperbolic, depending on frequency and doping. By considering the nonlocal response of BP for the first time, we have determined fundamental bounds in the confinement of BP plasmons. Importantly, results demonstrate that these bounds are looser than in graphene devices, allowing more tightly confined fields and stronger light-matter interactions in nanostructures. As expected, the quantum effects of nonlocality are anisotropic, with transitions between inductive and capacitive responses that depend on the in-plane direction and material properties. We have also demonstrated that nonlocality prevents plasmons from exhibiting true hyperbolic topology, contrary to previous predictions, and shown unusual nonlocality-induced topological transitions that leads to a broadband canalization regime. We envision a promising future for BP as a reconfigurable plasmonic platform whose electromagnetic properties can be further enhanced by nano-structuring or by realizing hybrid systems composed of BP, graphene, metals or other 2D materials, with exciting opportunities to realize a wide variety of optoelectronic devices, sensors, lenses and communications system at IR frequencies.

**Corresponding author**

*To whom correspondence should be addressed: alu@mail.utexas.edu

**Acknowledgements**

This work was supported by the Air Force Office of Scientific Research grant No. FA9550-13-1-0204, the Welch Foundation with grant No. F-1802, the Simons Foundation, and the National Science Foundation with grant No. ECCS-1406235. D. C. S. acknowledges financial support by the Spanish MECD under Grant FPU13-04982.